\documentstyle[aps,prb,amsfonts,preprint]{revtex}
\begin{document}

\draft
\tighten

\title{Is surface melting a surface phase transition?}
\author{Paul J. M. Bastiaansen\cite{paul} and Hubert J. F. Knops}
\address{Institute for Theoretical Physics,
University of Nijmegen, \\ Toernooiveld, 6525 ED Nijmegen, The Netherlands}

\maketitle

\begin{abstract}
Monte Carlo or Molecular Dynamics calculations of surfaces of Lennard-Jones
systems often indicate, apart from a gradual disordering of the surface
called surface melting, the presence of a phase transition at the surface,
but cannot determine the nature of the transition. In the present paper, we
provide for a link between the continuous Lennard-Jones system and a lattice
model. We apply the method for the (001) surface of a Lennard-Jones fcc
structure pertaining to Argon. The corresponding lattice model is a Body
Centered Solid on Solid model with an extended range of interaction,
showing in principle rough, flat and disordered flat phases. We observe that
entropy effects considerably lower the strength of the effective couplings
between the atoms. The Argon (001) face is shown to exhibit a phase
transition at $T=70.5 \pm 0.5$~K, and we identify this transition as
roughening. The roughening temperature is in good correspondence with
experimental results for Argon.
\end{abstract}


\section{Introduction and motivation}

Surface melting, in particular of simple Lennard-Jones systems like
Argon, is well understood as the wetting of the solid by its own
melt. A liquid-like layer appears between the solid and the vapor, and
the thickness of this layer increases with temperature.
Theoretically, the phenomenon is described with a Landau
theory\cite{Lipowsky82} and, more recently, using a Density Functional
approach.\cite{Lowen94}

It is clear that the atoms in the liquid-like layer between the bulk and
the vapor are influenced
by the presence of the underlying crystal. Therefore, the layer should
be regarded as a quasi-liquid exhibiting properties that are intermediate
between those of the solid and the bulk liquid. To decide upon surface
melting in Molecular Dynamics calculations, one usually considers the
behavior of an appropriate quantity like, e.g., the parallel integrated
density
\begin{equation}
   \rho(z) = \int\text{d}x\text{d}y \; \rho({\bf r}) ,
\end{equation}
where $z$ is the direction transversal to the interface, and
\begin{equation}
   \rho({\bf r}) = \sum_j \delta ({\bf r} - {\bf r}_j)
\end{equation}
with $j$ running over all particle positions ${\bf r}_j$. This $\rho(z)$
shows sharp peaks in the bulk, while in the case of surface melting the
peaks broaden when $z$ approaches the interface. The onset of the
quasi-liquid layer then can be defined, more or less arbitrarily, by a
suitable broadening of the peaks. In the case of complete surface
melting, the thickness $l$ of the quasi-liquid layer diverges as the
temperature approaches the triple point temperature $T_3$ of the bulk.
As follows from the wetting theory in the case of Lennard-Jones systems,
the thickness of the layer increases with temperature as a power law,
and this is experimentally confirmed.\cite{Zhu86}

The wetting theory concentrates on this parameter $\rho(z)$ but cannot give
information concerning the atomic
structure in the few top layers when the triple point $T_3$ is approached.
Specifically, one could ask whether the top layer does exhibit a genuine
surface phase transition at a temperature $T_c < T_3$. It is possible
that the phenomenon of surface melting is a gradual process of the
thickening of the quasi-liquid layer, with no surface phase transition at
all. But since the wetting theory fails to detect a genuine two-dimensional
transition, it could well be that surface melting is accompanied by a surface
phase transition which takes place
at a temperature lower than the triple point temperature. Such a
transition could either be a roughening transition, involving the
transversal degrees of freedom, or a two-dimensional `melting' transition
involving the in-plane degrees of freedom. The connection between surface
melting and possible surface phase transitions is as yet not understood.

One of the difficulties in addressing this question is the definition of
a suitable order parameter. Clearly, the parallel integrated density
cannot be an order parameter.

Van der Eerden {\it et al.}\cite{Eerden90,Eerden92,Eerden93} proposed the
surface shear modulus as an
order parameter for a surface phase transition accompanying surface
melting. In a Monte Carlo simulation of the (001) face of a Lennard-Jones
fcc structure, they found indications for such a transition, but, as is in
general the case with such calculations, no information regarding the nature
of this transition was obtained. It is this question we want to address in
this paper.

At the outset it should be clear that a `melting' transition associated
with the lateral degrees of freedom of the top layer cannot be a
two-dimensional melting transition of the Nelson-Halperin-Young
type.\cite{Halperin78,Nelson79,Young79,Strandburg88} This is because the
bulk under the top layer provides a substrate potential which is
commensurate to the top layer. Such a potential is
known\cite{Halperin78,Jose77} to be relevant in the renormalization group
sense. It is therefore more appropriate to adopt a lattice model,
with the lattice dictated by the substrate. Judging from the Monte Carlo
data from simulations on Lennard-Jones interfaces well below the triple
point,\cite{Eerden92,Eerden93,Broughton83b} neglecting overhangs and
vacancies is not a serious approximation. Accepting this approximation,
one arrives at a Solid On Solid (SOS) model.

It is by now well known that, if one allows for more than nearest
neighbor interactions, the phase diagram of these SOS models can be very
rich.\cite{Bernasconi93} Possible phase transitions include
roughening,\cite{Nolden87} preroughening,\cite{DenNijs87} and
deconstruction.\cite{Villain88} The preroughening and deconstruction
transitions are from a flat or reconstructed phase into a so-called
Disordered Flat (DOF) phase, in which the surface is disordered but remains
flat on average. Typically such a transition can occur when the next nearest
neighbor couplings become strong as compared to the nearest neighbor
couplings.

The couplings in the lattice model are effective couplings to be
calculated from the original continuous Lennard-Jones system. In the
context of lattice models, one usually just estimates the values of the
couplings, e.g. by counting the `broken bonds'. In deriving a lattice
model from the continuous Lennard-Jones system, however, one should
integrate the continuous degrees of freedom, thereby obtaining effective
couplings which typically contain a gain term due to the increased potential
energy (the broken bond), but also a loss term due to an increase of
entropy, since a particle in a cell adjacent to a vacant cell has more
freedom to move. This effect lowers in particular the strength
of the nearest neighbor coupling.

It is the goal of the present work to actually calculate the effective
couplings for a lattice model, pertaining to the (001) surface of a
Lennard-Jones fcc structure, and to analyze this model. In this way we can
determine the nature of a possible transition at this interface, which is
not feasible with Monte Carlo or Molecular Dynamics calculations. We use a
Lennard-Jones potential appropriate for Argon, calculate the effective
couplings of the corresponding lattice model by integrating the continuous
degrees of freedom of the Lennard-Jones system, and analyze the
resulting SOS model.
\\[2ex]
This paper is organized as follows. In Sec.~\ref{outline} we describe the
method used to arrive at the lattice model, starting from the Lennard-Jones
system. In Sec.~\ref{stepenergies} we carry out the calculations and present
the results. In Sec.~\ref{latticemodel} we give a description of the SOS
model that pertains to the (001) interface and present the phase diagram
of the model. We end with a conclusion. Several checks, validating the
approximations and testing the procedures, are addressed in the Appendix.

\section{Outline of the method}
\label{outline}

Deriving the lattice model from the continuous Lennard-Jones system
requires a number of steps. First we have to establish a cell description of
the surface. We notice from the simulations of Broughton and
Gilmer\cite{Broughton86} that
there is a considerable range of temperatures up to about 75~K where the top
layer of the (001) face shows already an irregular pattern while the atoms
in all other layers are close to their average positions and still behave as
bulk atoms. Note that the triple point of Argon found by Broughton and
Gilmer is at $T_3=82.7$~K (compared to the experimental
value\cite{Flubacher61} of $T_3 = 83.810$~K). Concentrating on a
temperature range up to $T=75$~K, it is a good approximation to neglect
correlations between the top layer and the bulk, thus treating the bulk in
a mean field manner. We use Monte Carlo simulations
to evaluate the average bulk lattice distance $a(T)$ as a function of
temperature. This distance $a(T)$ then defines the cell dimensions of the
surface cells. For the (001) face, these cells are rectangular blocks
with dimension $\frac{1}{2}\sqrt{2}\;a(T)$ centered around the average atom
position. The height of the cell is chosen such that the average potential
at the top of the cell is negligible.

As the substrate potential is caused by layers under the surface which, as
follows from the calculations of Broughton and Gilmer, behave as bulk
layers, we can calculate the substrate potential, present in the cells, in
a bulk simulation.

Having defined the cells with the substrate potential, we can evaluate the
effective (temperature dependent) couplings in the lattice model, arising
from the continuous Lennard-Jones system. Therefore we choose several
different surface configurations with some cells occupied and some empty,
and calculate the corresponding free energies by integrating the
continuous degrees of freedom of the atoms in their cells. Comparing this
free energy with the free energy of the fully occupied surface, we arrive at
a lattice model of the (001) surface, with the only degrees of freedom left
being discrete and describing whether or not an atom is present on its
site.

In this way one can arrive in principle at an exact representation of the
continuous system as a lattice model, but in practice one only considers
effective couplings extending over a limited range. Instead of sticking to
the language of atom-atom couplings, we will express the energy of a
configuration in terms of elementary step and kink configurations at the
surface. Limiting the range of the couplings to the next-nearest neighbor
distance, the four elementary step and kink configurations are those depicted
in
Fig.~\ref{vertices}. We calculate the free energies of nine different
surface configurations and express the free energies as good as possible in
terms of the vertex free energies $F_1$ to $F_4$. The accuracy of this match
gives an estimate of the error in neglecting couplings with a range
beyond the next-nearest neighbor distance.

The nine different configurations are chosen to fit on a strip of
$3\times\infty$ cells which is periodically repeated. This makes it possible
to find their surface free energies by a transfer matrix method.

The final step in the procedure is to analyze the resulting SOS
model. Actually the appropriate SOS model for a (001) face of an fcc
structure is the Body Centered Solid On Solid (BCSOS) model, which is
directly related to the six vertex model\cite{VanBeijeren77} with an
extended range of the interactions. A section of the phase diagram of this
model has recently been investigated by us\cite{Wij95} showing that the
inclusion of next nearest neighbor interactions indeed is capable of
stabilizing a DOF phase. We use the same method as described in this
reference to explore the occurrence of a two-dimensional phase transition
along the temperature path from 50~K up to 75~K in the phase diagram.

\section{Effective step energies}
\label{stepenergies}

We derive the BCSOS model from the continuous
Lennard-Jones system by calculating the (temperature dependent) effective
step energies $F_1$ to $F_4$ from Fig.~\ref{vertices}. The system under
study is a Lennard-Jones fcc structure pertaining to Argon. The precise
potential we use is the same as that used in Ref.~\onlinecite{Eerden92},
\begin{equation}
   V(r) = 4.569 \varepsilon
   \left[ \left( \frac{r}{\sigma} \right) ^ {-12} -
           \left( \frac{r}{\sigma} \right) ^ {-6 }
   \right]
   \exp\left( \frac{0.25\sigma}{r-2.5\sigma} \right)
   \text{~~for~~}
   r < 2.5 \sigma,
   \label{LJ}
\end{equation}
and vanishes for $r \geq 2.5\sigma$. To model Argon, the following
values are used:
\begin{eqnarray}
   \frac{\varepsilon}{k} &= 119.8~\text{K} \\
   \sigma                &= 0.33~\text{nm}.
\end{eqnarray}
These values define the scale of our calculations, and can be used to
transfer the numerical results in terms of reduced units. As our
calculations particularly pertain to Argon, we choose to express the
results in terms of SI units.

Bulk simulations, used for establishing the cell description of the surface,
are treated in Sec.~\ref{bulksim}. In Sec.~\ref{transfermatrix}, we use a
transfer matrix method to calculate the effective step free energies.
Results are presented in Sec.~\ref{results}

\subsection{Bulk simulations}
\label{bulksim}
This section deals with computing the effective substrate potential to
which the surface atoms are subject. The substrate potential arises from
the bulk. The main assumption regarding this section is that there are
only two types of atoms: bulk atoms and surface atoms. Surface atoms are
those in the top layer; the remainder are bulk atoms. The assumption is
that atoms close to the surface are not affected by the surface; the
substrate potential in the surface layer thus is dictated exclusively by the
behavior of the bulk and can be determined in a bulk simulation.

Therefore we define a bulk system, consisting of 972 atoms in an fcc
configuration with periodic boundary conditions in every direction. It is
oriented such that a (001) layer of the configuration lies horizontally. We
choose one (001) layer of atoms from this configuration. The substrate
potential at a certain point in this layer now arises from all atoms
{\it under}\/ this layer.

Extracting the desired potential pattern consists of three steps: first
we have to compute the equilibrium volumes of the system at several
temperatures and fixed external pressure. In a next simulation, we fix this
equilibrium volume and compute the average positions of each of the atoms.
These average positions together with the equilibrium lattice parameter
determines the cells corresponding to each of the atoms. These
cells are rectangular blocks, and the center of the cell is the averaged
position of the corresponding atom. We can now rerun the
second simulation (i.e.\ generating the same configurations), but now
with the average atom positions known beforehand. In the cells, we
compute the average substrate potential pattern arising from all
atoms {\it under} the layer where the cell is part of. As, due to
translational invariance, all cells are equivalent, we can average all
measured potential patterns, thus giving rise to one average pattern in
a cell. This pattern is the desired substrate potential.

We do calculations in the temperature range of 50 to 75~K, still well below
the triple point of Argon. We will use an external pressure $P=0$ for our
simulations, which is, strictly spoken, not correct, as we do have to
consider the system in equilibrium with its vapor. The vapor pressure,
however, will be very low, and the properties of a solid are relatively
insensitive to pressure. We regard choosing $P=0$ as a good approximation.

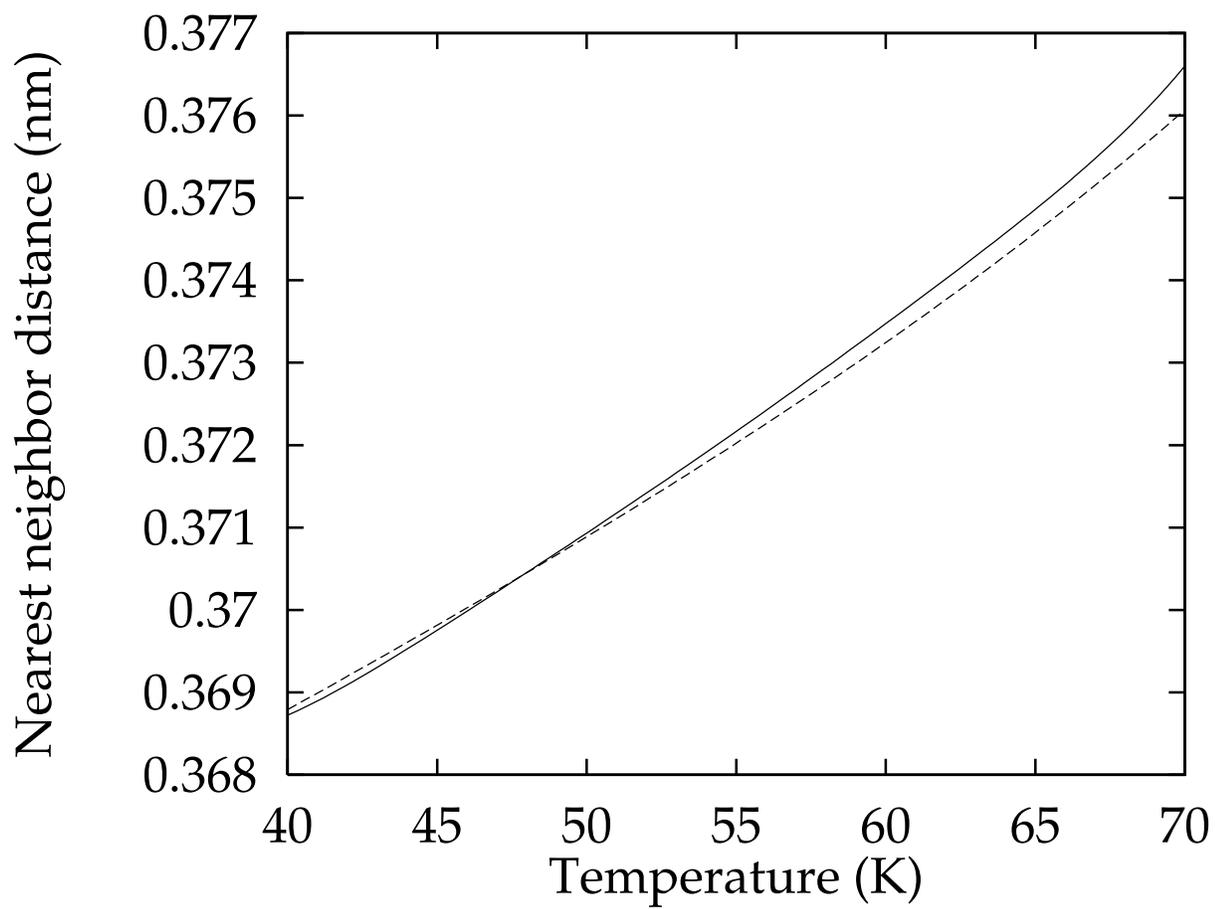
\begin{figure}
   \caption{The nearest neighbor distance of the Argon fcc crystal
   against temperature. The solid line is a polynomial fit of our
   results, the dashed line is that of Broughton and Gilmer. Note the
   $y$-scale.}
   \label{nndistances}
\end{figure}

The results of the volume calculations, expressed as nearest neighbor
distances and compared to those of Broughton and Gilmer,\cite{Broughton83a}
are shown in Fig.~\ref{nndistances} and show a close correspondence.
The thermal cubic coefficient of expansion $\alpha$ is given by
\begin{equation}
   \alpha = \frac{1}{V}
   \left(
      \frac{\partial V}{\partial T}
   \right)_P.
\end{equation}
We find at $T=60$~K a value of
$\alpha = 2.15\times 10^{-3}\text{~K}^{-1}$, as compared to the result of
Broughton and Gilmer\cite{Broughton83a} calculated from their polynomial
fit of the nearest neighbor distance,
$\alpha = 2.04\times 10^{-3}\text{~K}^{-1}$, and the result of Van der
Eerden {\it et al.},\cite{Eerden93}
$\alpha = 1.95\times 10^{-3}\text{~K}^{-1}$.
Note that Broughton and Gilmer use a slightly different Lennard-Jones
potential.

Having carried out the volume measurements, we can start measuring the
potential pattern. We perform the simulations with a fixed volume $V$,
and fix this volume at its equilibrium value as described above.
In the first simulation, we calculate the average position of all
atoms to determine the centers of the cells. During the simulation, we
store the generated configurations which we used for measurements, in order
to use them again in the next simulation.

Then we define in every cell a fine grid consisting of
$21 \times 21 \times 21$ points, at which we measure the average
potential. In course of our simulations, it turned
out that we had to extend the grid above the cell, because we wanted to
measure the potential up to a height where it becomes negligible. The
defined grid therefore has a height of 3/2 of the height of the bulk cell.
In the simulation, we use the already stored configurations. We measure
the substrate potential in each of the 9261 points of the grid, and average
over all cells. During the simulation, we have to check whether the cell
picture makes sense. It does, provided the atoms remain in their cell
during the simulation. We check this for all atoms, and it turns out that
not a single atom moves out of its cell.

As expected, the fluctuations in the potential are relatively large,
which means that we have to simulate long. We used 800 measurements,
with 3000 generated configurations between each measurement.

\subsection{Transfer matrix calculations}
\label{transfermatrix}

The first step in our procedure, the determination of the substrate potential
pattern, has now been carried out. We will use this potential pattern
as a mean field, to which the surface atoms are subject. However, we
have to check whether this mean field approximation makes sense, and
indeed will do so using different methods. The checks will be treated
in the Appendix.

Our task is to integrate the continuous degrees of freedom of the
surface atoms in their cells. We will do this using a transfer matrix
calculation, which, in principle, gives exact results. We have, however,
to discretize as a continuous integration is not possible. If we choose a
surface in which one row consists of $N$ cells with $L$ grid points per cell,
the dimension of the transfer matrix is $L^N \times L^N$. The boundary
conditions are periodic: the row is closed at the ends.

The substrate potential pattern consists of 9261 points, which means
that now $L=9261$. This yields a much too large dimension of the
transfer matrix. Therefore, we have to reduce drastically the number of
points $L$. We choose $L=25$ and $N=3$, yielding a linear dimension of
the transfer matrix $L^N=15625$. We will distribute the 25 points in the
cell as efficiently as possible, by choosing the grid points unevenly
distributed in the cell and making the pattern two-dimensional. Both
approximations, reducing the number of points and making the grid
two-dimensional, will be validated.

The Hamiltonian of our surface system is
\begin{equation}
   H = \sum_i V_{\text{subs}}( {\bf r}_i ) +
       \sum_{i,j} V_{\text{LJ}}( |{\bf r}_i - {\bf r}_j| ),
\end{equation}
where the indices run over all particles, $V_{\text{subs}}$ is the
substrate potential and $V_{\text{LJ}}$ is the Lennard-Jones potential
from Eq.~(\ref{LJ}). $V_{\text{LJ}}$ is cut off such that only nearest and
next nearest neighbors interact; otherwise the transfer matrix dimension
increases.

Making the grid two-dimensional means that we ignore the dependence of
$V_{\text{LJ}}( |{\bf r}_i - {\bf r}_j| )$ on $z_i$ and $z_j$. The $z$
integration can then be done directly in the partition function. This is
done by just adding all potential patterns with different $z$
coordinates, and multiplying the corresponding Boltzmann weights by the
height of the grid.

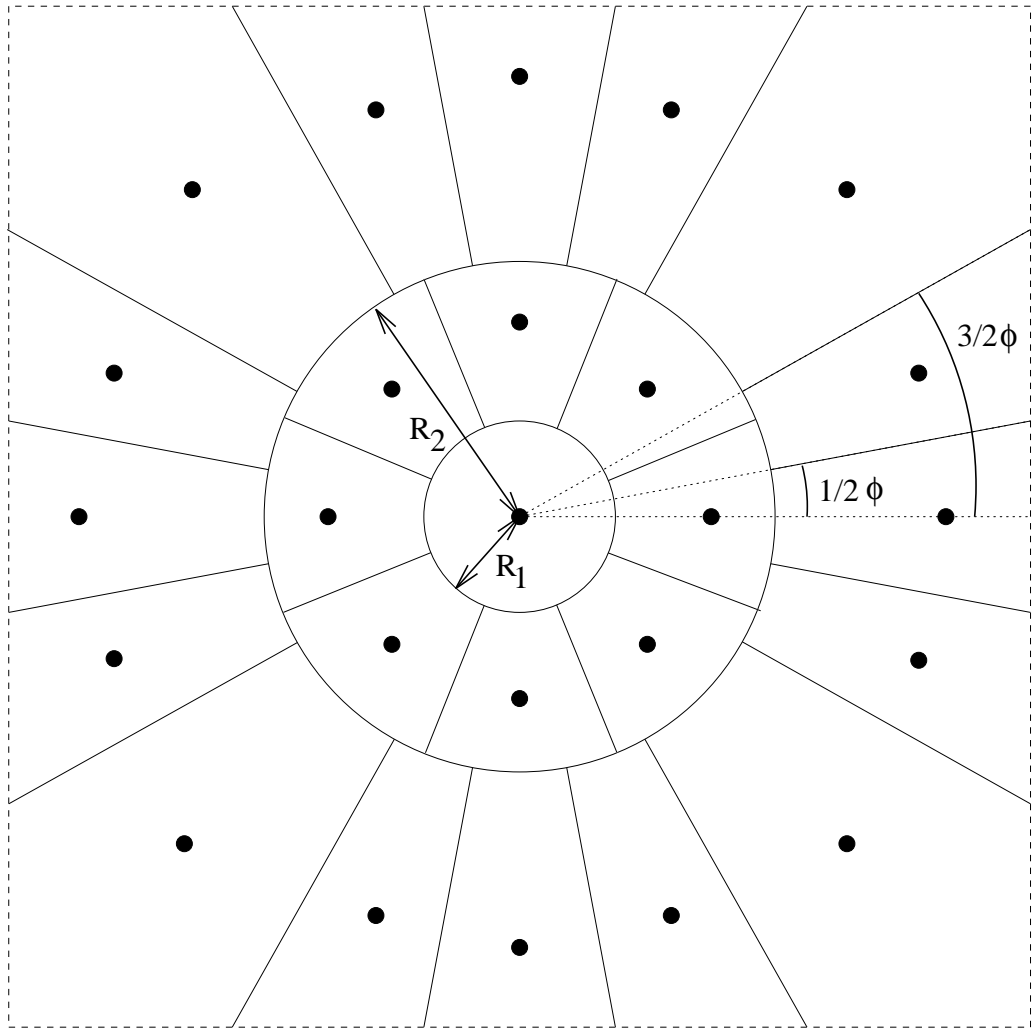
\begin{figure}
   \caption{Distribution of points in a cell. There are 25 points in the
   cell, each one in its corresponding domain. The potential in each of
   the domains is averaged, its Boltzmann weight is calculated and
   multiplied by the area of the domain. The points are in the
   middle (the `center of mass') of the domain, and the domains are
   uniquely defined according to the parameters $R_1$, $R_2$ and $\phi$.
   Note that symmetry is that of a square. The parameters are fixed such that
   the Boltzmann weights multiplied by the corresponding area are all equal,
   in order to obtain a distribution which is as efficient as possible.}
   \label{pointdistr}
\end{figure}

Then we choose the $L=25$ grid points as efficiently as possible. The
idea is as follows: the lower the Boltzmann weight in a certain point of
the cell, the less relevant is the corresponding area in the partition
function, so the distribution can be less dense in that area. On the
other hand, points have to be dense in those areas where the Boltzmann
weights are high. Grosso modo this means that the distribution of points
should be more dense in the center of the cell and less dense towards
the edges. We choose the location of the points according to
Fig.~\ref{pointdistr}. The figure depicts the two-dimensional cell with
25 points, each point being in its corresponding domain. The points are in
the middle (the `center of mass') of their domain. The domains are defined
by the three variables $R_1$, $R_2$ and $\phi$.

We choose the grid points as in the figure, and we assign a Boltzmann
weight to each of the domains. Consider one domain with the original
(dense) grid points being in it. To obtain the correct Boltzmann weight,
we first determine the average potential $V$ in that domain, calculate the
corresponding Boltzmann weight $W=\exp (-V)$ and multiply this with the
area of the domain. This is the Boltzmann weight we are going to use in
the transfer matrix calculation. To obtain the most efficient
distribution of points, we fix the variables $R_1$, $R_2$ and $\phi$
such that the Boltzmann weights multiplied by their corresponding area,
are (almost) equal to each other. A lower Boltzmann weight then
automatically corresponds to a larger domain, and the resulting weights
are equally relevant. Theoretically, they are all exactly equal so we
could leave them out, but in practice we do not succeed in fixing the
parameters such that the weights are exactly equal, so we choose to
leave them where they are.

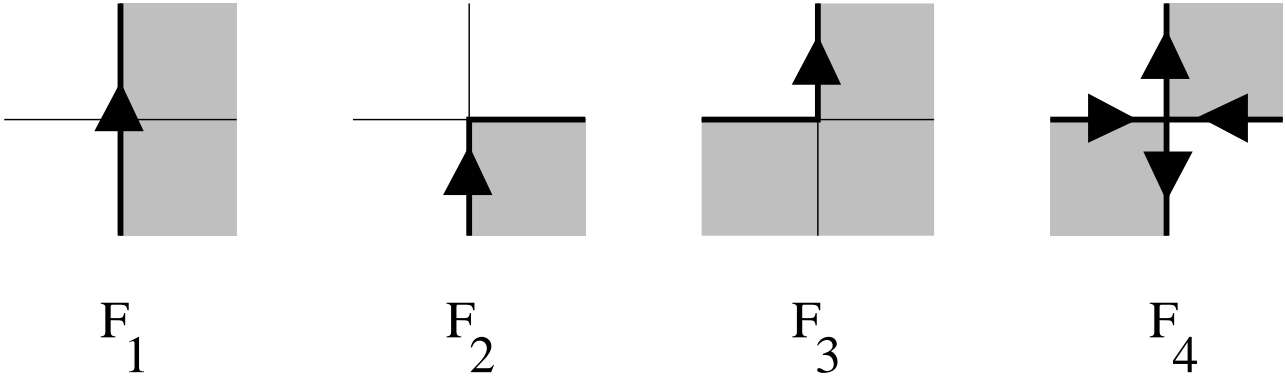
\begin{figure}
   \caption{Four different step configurations at a vertex. Shaded cells
   are occupied, white cells are empty. Thick lines are the oriented
   steps a the vertex configuration.
   $F_j$ denotes the free energy. Note that the configurations are
   rotationally invariant, so $F_1$ represents four different
   configurations. The empty configuration, without step, is not shown
   and has a free energy $F=0$.}
   \label{vertices}
\end{figure}

The calculated coordinates of the 25 grid points, together with the
associated Boltzmann weights, are used as input for the transfer matrix
calculations. We calculate the free energy of several different surface
configurations, each consisting of occupied and empty cells. We choose them
such as to have much variety in the configurations, and we choose them
overcomplete for determining the vertex free energies $F_1$ to $F_4$ from
Fig.~\ref{vertices}.
The configurations we choose are depicted in Fig.~\ref{confs}, where the
pictures are understood to extend to infinity at the right and left ends.
The `rows' of the transfer matrix are depicted vertically; the direction
of transfer is horizontal. We have to
define a different transfer matrix for each different pair of rows;
transferring from a row with, say, three occupied cells to a row with
two occupied cells corresponds to a transfer matrix with dimension
$25^3 \times 25^2$.

Let us define the different transfer matrices by, for example,
$T(110|111)$, where the digits refer to the cells. A digit 0 indicates
that the corresponding cell is empty, a 1 that it is occupied. So
$T(110|111)$ is the transfer matrix between a left row having its lower
two cells occupied, and a right row having all cells occupied. Note that
none of the transfer matrices is symmetric and that most of them are not
even square matrices.

First we calculate the largest eigenvalue $\lambda_{\text{max}}$ and the
corresponding left and right eigenvectors $\langle\psi|$ and $|\psi\rangle$
of the `full' transfer matrix $T(111|111)$. The free energy of one particle
in a fully occupied surface is $F=-\frac{1}{3}\ln \lambda_{\text{max}}$.
Now the partition function of, say, configuration number~9 in
Fig.~\ref{confs} is
\begin{equation}
   Z = \frac{1}{\lambda_{\text{max}}^2} \;
   \langle\psi| \; T(111|101) \: T(101|010)  \: T(010|111) \; |\psi\rangle .
\end{equation}
The eigenvectors are normalized as $\langle\psi|\psi\rangle=1$. Note that
we divide by $\lambda_{\text{max}}^2$, because we want to subtract the free
energy of a fully occupied surface consisting of precisely as many atoms as
configuration number 9. Similar expressions are used for the other
surface configurations.

By calculating $F=-\ln Z$ we now have the free energies of all different
surface configurations. We want to express those free energies in terms
of step configurations at the surface. The step configurations we use
are those possible at a vertex (the meeting point of four cells) and are
depicted in Fig.~\ref{vertices}. The fifth possible vertex is the vertex
with no steps; it carries free energy 0. The nine free energies of the
surface configurations of Fig.~\ref{confs} are redundant for determining
the values of the free energies $F_1$ to $F_4$. We calculate $F_j$ using
a best fit method. Re-expressing the free energies of the nine surface
configurations in terms of these vertex free energies indicates the
quality of the fit.

Ideally, the $F_j$ obey the following equation
\begin{equation}
   \sum_{j=1}^{4} T_{ij} F_j = A_i
   \label{fit}
\end{equation}
where $A_i, i=1\ldots 9$ is the free energy of surface configuration
number $i$ and $T_{ij}$ expresses the number of times vertex $j$ appears
in configuration number $i$, that is
\begin{equation}
   \tilde{T} = \pmatrix{ 6&8&8&6&6&2&4&4&2 \cr
                         0&2&2&4&0&4&4&2&0 \cr
                         0&2&2&0&4&4&4&4&4 \cr
                         0&0&0&0&0&0&0&1&2 } .
\end{equation}
As argued before, the nine surface configurations are redundant for
determining the free energies $F_j$, so Eq.~(\ref{fit}) will have no
solution. We will determine the $F_j$ from this equation by introducing an
error vector $\varepsilon_i, i=1\ldots 9$ by
\begin{equation}
   \sum_{j=1}^{4} \frac{T_{ij}F_j}{A_i} = 1 + \varepsilon_i .
\end{equation}
The components of $\varepsilon_i$ describe the relative error of
expressing the free energy of configuration number $i$ in terms of the
vertex free energies $F_j$. Minimizing $|\bbox{\varepsilon}|^2$ with respect
to the vertex energies by solving
\begin{equation}
   \frac{\partial |\bbox{\varepsilon}|^2}{\partial F_j} = 0
\end{equation}
yields the best values for $F_j$.

\subsection{Results}
\label{results}

The resulting values for $F_1$ to $F_4$ are plotted in
Fig.~\ref{freeenergies}. It turns out that expressing the surface free
energies in terms of only these elementary configurations works well;
the average percentual difference between the actual free energy and the
free energy in terms of the $F_j$ is about 1.2\%, the maximum difference
is 2.5\%.

In lattice models, one usually estimates the interaction parameters by
the energy of a broken bond. This energy is, in our case, the value of the
Lennard-Jones potential at the equilibrium distance. One thus neglects
the entropy gain that corresponds to the increased freedom of the moving
atom. To see the strength of this effect, we choose one of the surface
configurations, say configuration number~7, and plot its energy
calculated from the broken bonds together with the actual free energy
which results from our transfer matrix calculations. The plot is shown in
Fig.~\ref{conf7}. There turns out to be a considerable difference in
the actual free energy and the energy calculated by counting the broken
bonds. This difference is precisely the entropy which the atoms gain by
moving in the cell, and, as expected, it increases with temperature.

To visualize the quality of the fit method described above, we also plot
(in Fig.~\ref{conf7}) the free energy of the configuration, but now
re-expressed in terms of the vertex free energies $F_j$ from
Fig.~\ref{vertices}. We see from the figure that expressing the
free energy of the configuration in terms of the vertex energies is
appropriate. We conclude that it is sufficient to consider only the vertex
configurations 1 to 4, and that we do not need to consider more complex
step configurations.

Finally, we want to check an hypothesis. We expected that the decrease
in free energy of the surface with increasing temperature is mainly due
to the increasing lattice parameter, and that it has little to do with
the changing substrate potential pattern. Cell dimensions increase with
temperature, which means that the atoms gain freedom and therefore
entropy, while the potential pattern flattens a little but remains more
or less the same. We can check this hypothesis by scaling the potential
pattern at
$T=50$~K to the cell dimensions at $T=70$~K, and calculating the free
energies of the surface configurations. The results are tabulated in
table~\ref{checktable}. We conclude that the increase of free energy
with temperature can be explained for 98\% by the
increasing cell dimensions, and for roughly 2\% by the flattening of the
potential pattern. This confirms our hypothesis.

\section{The lattice model}
\label{latticemodel}

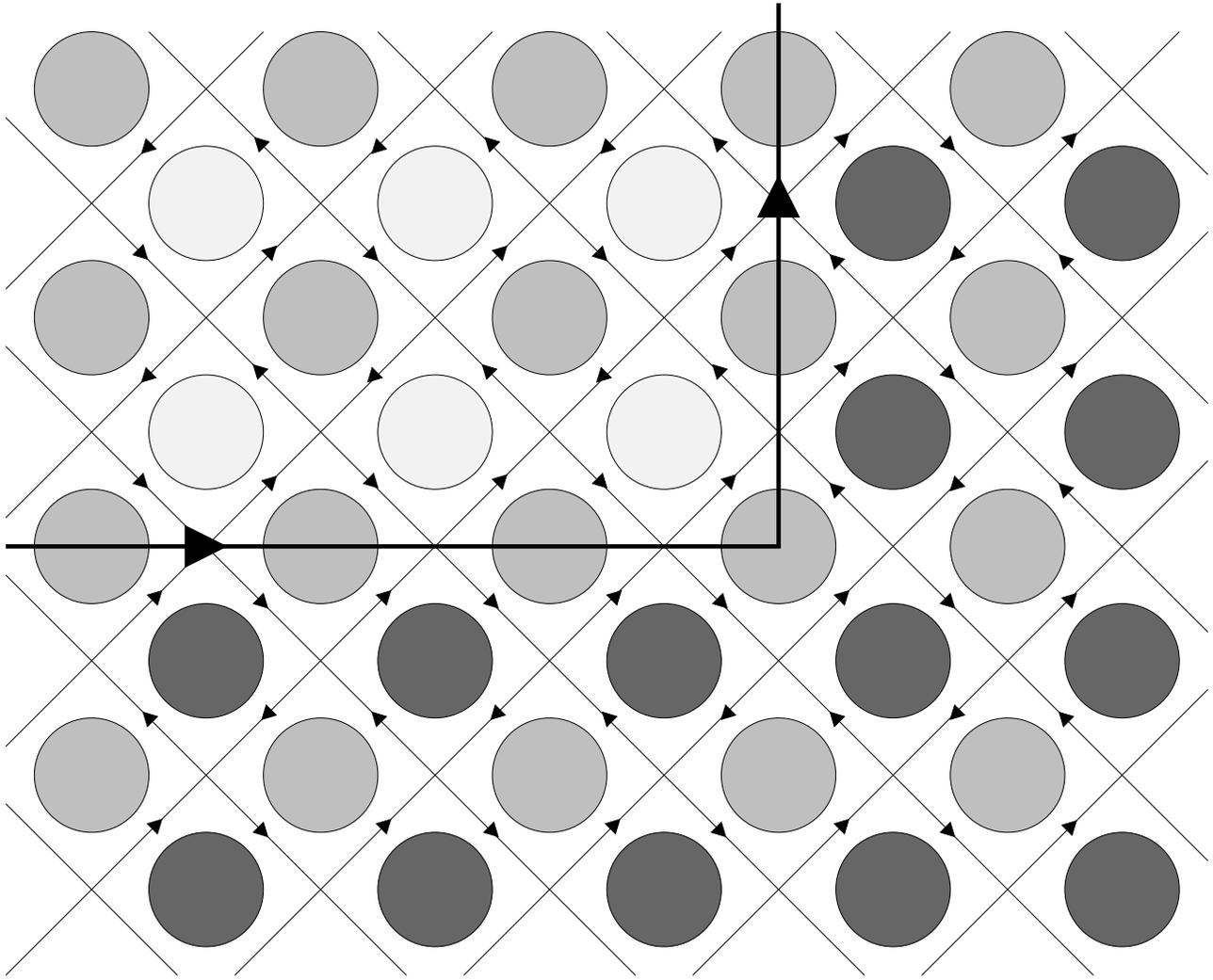
\begin{figure}[p]
   \caption{Typical surface configuration for a (001) surface of an fcc
   crystal. The circles are the atoms, darker circles representing atoms
   that are higher on the surface. Neighboring atoms always differ
   half a lattice distance in height. The six vertex lattice is depicted
   with thin, solid lines and small arrows. The thick solid line is a step
   on the surface, and indicates a height difference of one lattice
   distance $a$ between next nearest neighbors.}
   \label{surface}
\end{figure}

We will now briefly discuss the lattice model following from our
calculations. The model pertains to the (001) surface of an fcc crystal,
which means that neighboring atoms always differ $\pm\frac{1}{2}a$ in
height. A typical surface configuration is depicted in Fig.~\ref{surface}.
The resulting lattice model is a BCSOS model, which can be mapped on a six
vertex model.\cite{VanBeijeren77}

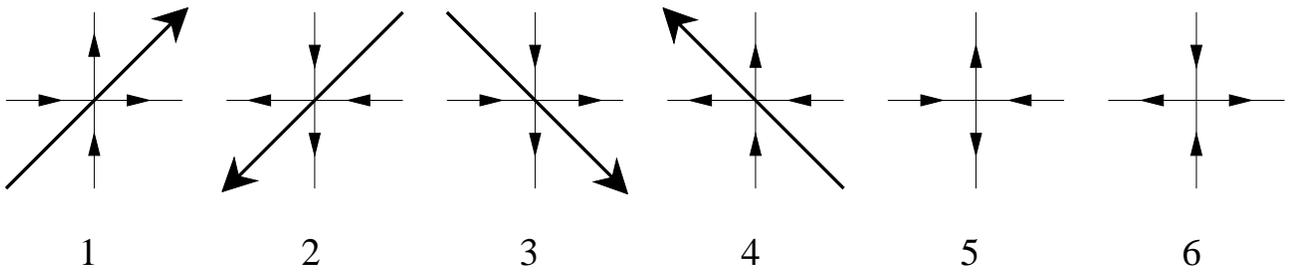
\begin{figure}[p]
   \caption{The six possible arrow combinations at a vertex of the six
   vertex model. The six vertex model can be mapped onto a surface model
   by assigning heights to each of the lattice sites. Looking in the
   direction of the arrow, the higher atom is on the right. The thick
   arrows denote height differences between next nearest neighboring
   atoms, and form a closed, oriented loop on the surface. The loop
   corresponds to a step.}
   \label{6vertex}
\end{figure}

Consider a square lattice, where an
arrow is placed on each of the bonds satisfying the so-called ice rule,
which states that the number of incoming arrows at each vertex equals
the number of outgoing arrows. The six possible arrow combinations at a
vertex are depicted in Fig.~\ref{6vertex}. The rule for assigning heights
to the lattice sites is that, when looking in the direction of an arrow,
the higher atom is on the right. Fixing then the height of one of the sites,
a vertex configuration is uniquely mapped onto a surface height
configuration. The ice rule guarantees that the mapping is single valued.

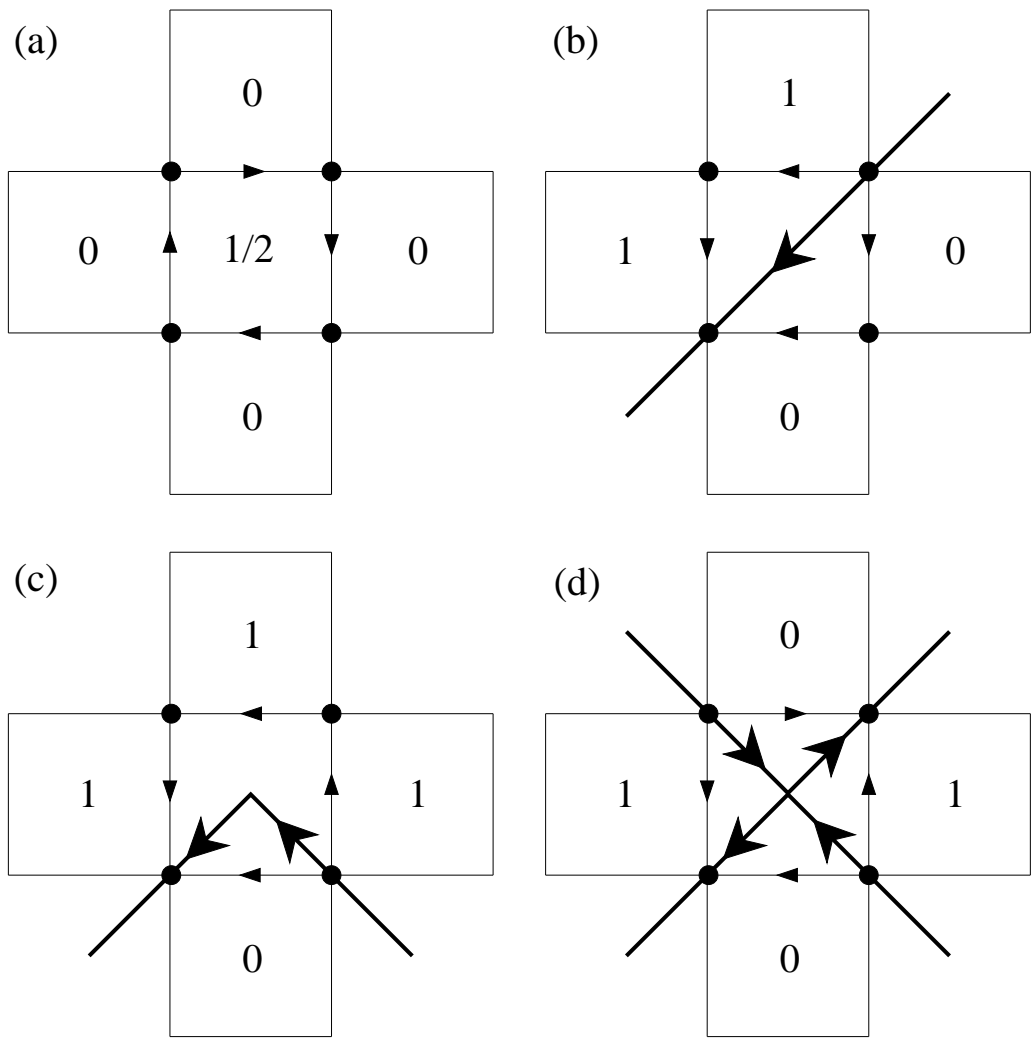
\begin{figure}[p]
   \caption{Different elementary surface configurations. Compare
   Fig.~\protect\ref{surface}, and note that the lattice is tilted over
   45$^\circ$. The thin solid lines make up the six vertex lattice, each
   bond carrying an arrow and indicating a height difference
   $\pm\frac{1}{2}a$ between adjacent atoms. Heights, in terms of the
   lattice parameter $a$, are indicated on the sites. The interaction is
   defined on the central side. For convenience also the adjacent sites
   are depicted. The four vertices represented as $\protect\large\bullet$
   take part in the interaction, but only insofar their arrows are depicted.
   Thick lines are steps on the surface, indicating a height difference
   between next nearest neighbor atoms. (a) is the flat configuration
   carrying no step, (b) is a straight step, (c) is a corner and (d) is a
   crossing. The free energy of (a) is 0, of (b) is $F_1$, of (c) is
   $(F_2+F_3)/2$, and of (d) is $F_4$. Note that we do not distinguish
   between inside and outside corners.}
   \label{loops}
\end{figure}

The six vertex model in its original formulation assigns an energy to each
of the six vertices, and has been completely solved.\cite{Baxter82} In
our case, using the configurations and energies of Fig.~\ref{vertices}, we
have to take into account interactions between the vertices, which
means
that there is no exact solution of our model present. The energies which
we assign to vertex configurations are visualized in Fig.~\ref{loops},
and result in a four-vertex interaction. Compare this figure to
Fig.~\ref{vertices}.

The model can be analyzed with transfer matrix calculations and standard
analysis of Finite Size Scaling.\cite{Nightinale82} In another
paper,\cite{Wij95} we analyze a section of the phase diagram of this
BCSOS model and show that it contains in particular both a rough and
a DOF phase. For a detailed account of the calculations we refer to this
reference. Here, we apply the same method to determine the behavior of the
Argon (001)
surface, following from the values of the $F_j$ in Fig.~\ref{freeenergies},
but average the free energies $F_2$ and $F_3$ applying to inside and outside
corners. Inequality of $F_2$ and $F_3$ breaks the particle-hole symmetry and
cannot be present for a two body potential. The broken symmetry is present
here because of entropy effects, which introduce effective many-body
couplings. The effect will be to smear out the transition\cite{DenNijs90}
but we neglect the difference here and use $(F_2+F_3)/2$.

Our Finite Size Scaling analysis shows that the model exhibits a
roughening transition at $T=70.5 \pm 0.5$~K. Below this temperature, the
surface is flat on average, with only now and then islands bounded
by up or down steps. At higher temperatures, these islands start to
grow, eventually diverging at the roughening temperature, where the step
free energy vanishes.

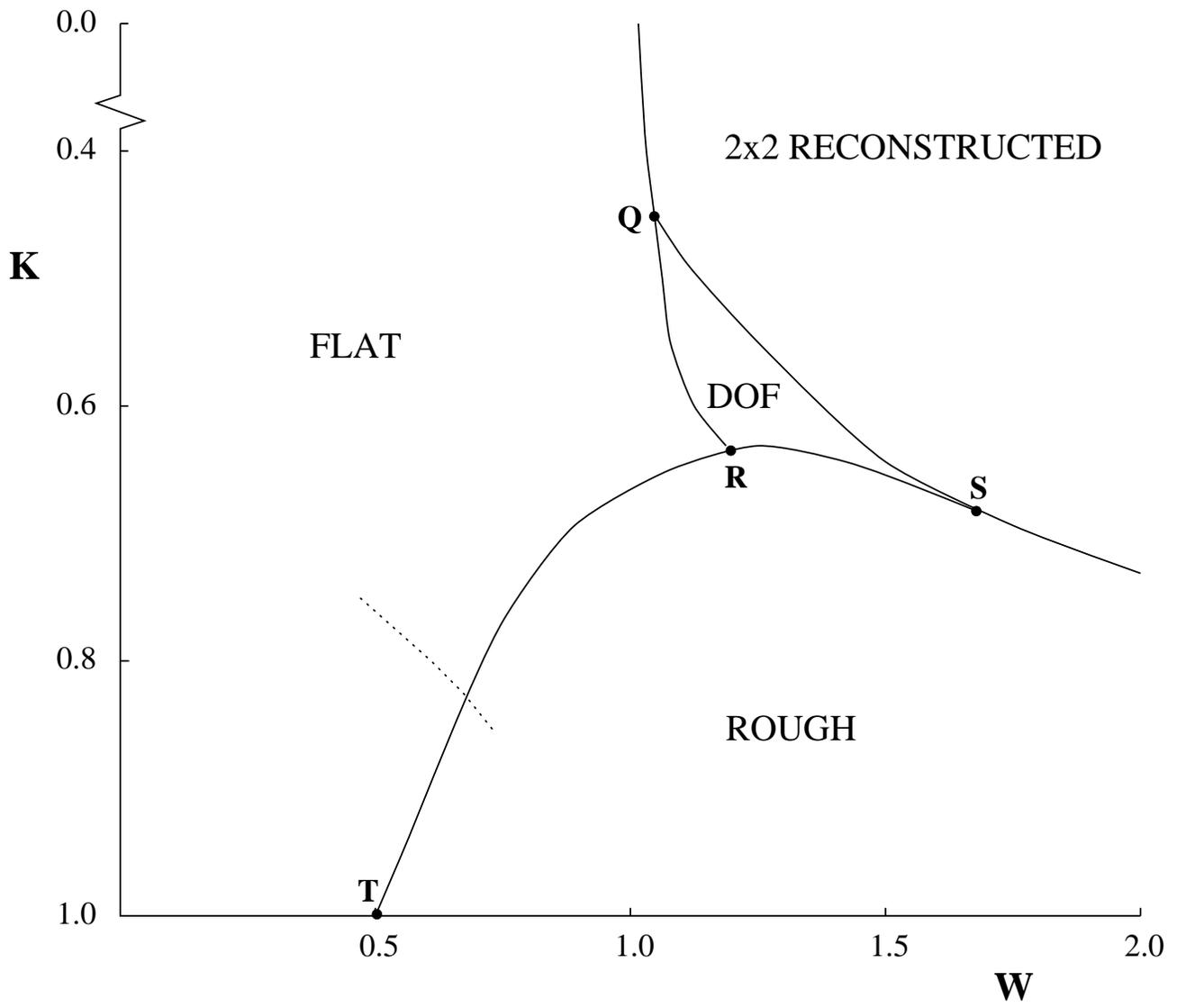
\begin{figure}[p]
   \caption{The phase diagram of the BCSOS model. The parameters $W$ and
   $K$ are the Boltzmann weights assigned to a broken next nearest neighbor
   bond and a broken third nearest neighbor bond respectively. The model
   exhibits flat, rough, Disordered Flat (DOF) and reconstructed phases
   as indicated. The dotted line roughly corresponds to the path the
   Agorn (001) surface is following as a function of temperature.}
   \label{phase}
\end{figure}

In the more general analysis of the phase diagram of this BCSOS model,
we assigned a Boltzmann weight $W$ to a broken next nearest neighbor bond
and a weight $K$ to a broken third nearest neighbor bond. Note that
nearest neighbors always differ $\pm\frac{1}{2}a$ in height. The phase
diagram of the model is depicted in Fig.~\ref{phase}. In terms of $W$ and
$K$, the $F_j$ would be
\begin{eqnarray}
   \beta F_1 &=& -\ln( WK^2 ) \\
   \beta F_2 &=& -\ln( WK   ) \\
   \beta F_3 &=& -\ln( WK   ) \\
   \beta F_4 &=& -\ln( W^2  ) .
\end{eqnarray}
The values of $F_j$ that follow from our calculations of course can not
be exactly expressed in terms of $W$ and $K$; we would need a higher
dimensional parameter space. However, the phase diagram gives an
indication of where the Argon (001) surface is located with respect to the
DOF and the rough phase. The dotted line in the phase diagram roughly
corresponds to this location.
\\[2ex]
We stress that in this way we are actually able to predict a surface
phase transition of a continuous Lennard-Jones system, and to identify
this transition as roughening, whereas determination of the nature of
the transition is not possible in direct Monte Carlo or Molecular Dynamics
calculations on the continuous system.

Zhu and Dash\cite{Zhu86} performed heat capacity measurements on thick
Argon films adsorbed on graphite. They observed, apart from surface
melting, weak, rounded anomalies at $T=68$~K, which they
identified tentatively as roughening. Roughening is accompanied by a small
peak in the heat capacity, which lies somewhat below the actual roughening
temperature $T_{\text{R}}$. Our value $T_{\text{R}} =70.5 \pm 0.5$~K is
in good agreement with the peak in the heat capacity at $T=68$~K.

Van der Eerden {\it et al.}\cite{Eerden93} performed Monte Carlo simulations
on the (001) face of a Lennard-Jones fcc structure, and observed the
vanishing of the surface shear modulus at $T=64\pm 2$~K. Regarding the
difficulties in determining this temperature, due to the fact that the
way in which the shear modulus vanishes is not known, and judging from
their figures, we find the roughening temperature $T_{\text{R}}$ as
found by us in correspondence with their findings.

\section{Conclusions}

We studied the (001) face of a Lennard-Jones fcc structure, particularly
pertaining to Argon. Direct Monte Carlo or Molecular Dynamics
simulations of such systems do indicate the presence of surface phase
transitions, but cannot unambiguously determine the nature of such a
transition; the appropriate way to do so is to use a discrete lattice
model. Therefore we provided for a link between the continuous
Lennard-Jones system and the corresponding lattice model, which is, in
this case, a BCSOS model. We describe a method to calculate the
effective interaction constants in the lattice model, arising from the
Lennard-Jones interaction, and carried out the calculations for (001)
surface. We observed that entropy effects, arising from the gain in
freedom of an atom adjacent to a vacancy, considerably lower the
interaction constants. The BCSOS model is shown to exhibit a roughening
transition at $T=70.5 \pm 0.5$~K, which is in good agreement with
experimental results for Argon.

Several checks on the approximations used for the calculations are dealt
with in the Appendix. We note that the approximations allow for an
estimate of their accuracy. In this way one can, when more accurate
numerical data are required, give a quantitative estimate of the effect
of the approximations and build in corrections.

The method described in this paper can be applied to other systems and
other surfaces as well. For the (110) surface of an fcc structure, the
same BCSOS model applies, albeit with different values of the couplings.
We stress that the phase diagram of the BCSOS model shows, apart from a
flat and a rough phase, also a Disordered Flat phase. It may be possible,
applying the method for the (110) surface, to determine a preroughening
transition from the flat into the Disordered Flat phase.
\\[2ex]
In the case of the (001) face of a fcc Lennard-Jones system,
surface melting, a disordering of the surface layers judged from
correlation functions, layer occupation etc., is shown to be accompanied
by a genuine surface phase transition, which is a roughening transition
in this particular case. A precise, atomic-scale definition of surface
melting is required to further examine the interplay between the
disordering process and the possibility of surface phase transitions.

\begin{acknowledgments}
We thank Jan van der Eerden for discussions and for providing the program
{\sc simlib}, used to carry out the Monte Carlo simulations.
\end{acknowledgments}

\appendix
\section*{}

Several additional calculations, validating the approximations we made,
have been performed and will be treated in this Appendix. The most
important approximations arise from the transition from three
dimensional to two dimensional, and from the reduction of the number of
points in a cell.

\subsection{Free energy of the bulk}

First we want a general check on all approximations. The
method we followed for calculating the surface free energies, including
all approximations, can also be used for calculating the free energy of
the bulk and consequently the vapor pressure, which can be compared with
experimental values for Argon. We do this as follows. Consider
a (001) layer in the bulk. This layer is subject to two substrate
potential patterns as calculated in Sec.~\ref{bulksim}; one from below,
and one from above. These two potential patterns are exactly those we
calculated already. Using those two patterns as the
potential pattern in a bulk (001) layer, we can apply the transfer
matrix to obtain the bulk free energy. The bulk free energy
is then obtained using all approximations we used for calculating the surface
free energy, that is, treating the potential pattern in a mean field
approximation, treating the cells as two-dimensional and drastically
reducing the number of grid points in the cell.

The full partition function of the bulk system consisting of $N$ atoms is
\begin{equation}
   Z_{\text{bulk}} =
      \frac{1}{h^{3N}}
      \int d^3\bbox{p}_j \exp\left( -\frac{\beta\bbox{p}^2}{2m} \right)
      \;Z_{\text{conf}},
      \label{partfunc}
\end{equation}
where $h$ is Planck's constant. Note that we do not need correct
Boltzmann counting; the particles can be identified by their cells.
The first part of this expression is the trivial kinetic part of the
energy, the second part is the configurational part which equals
\begin{equation}
   Z_{\text{conf}} = \lambda_{\text{max}}^{N/3},
\end{equation}
where $\lambda_{\text{max}}$ is the largest eigenvalue of the transfer
matrix with the bulk potential pattern. Note that this eigenvector pertains
two a row with three atoms in it; hence the factor $N/3$. Note also that
$\lambda_{\text{max}}^{1/3}$ has the dimension of a volume.

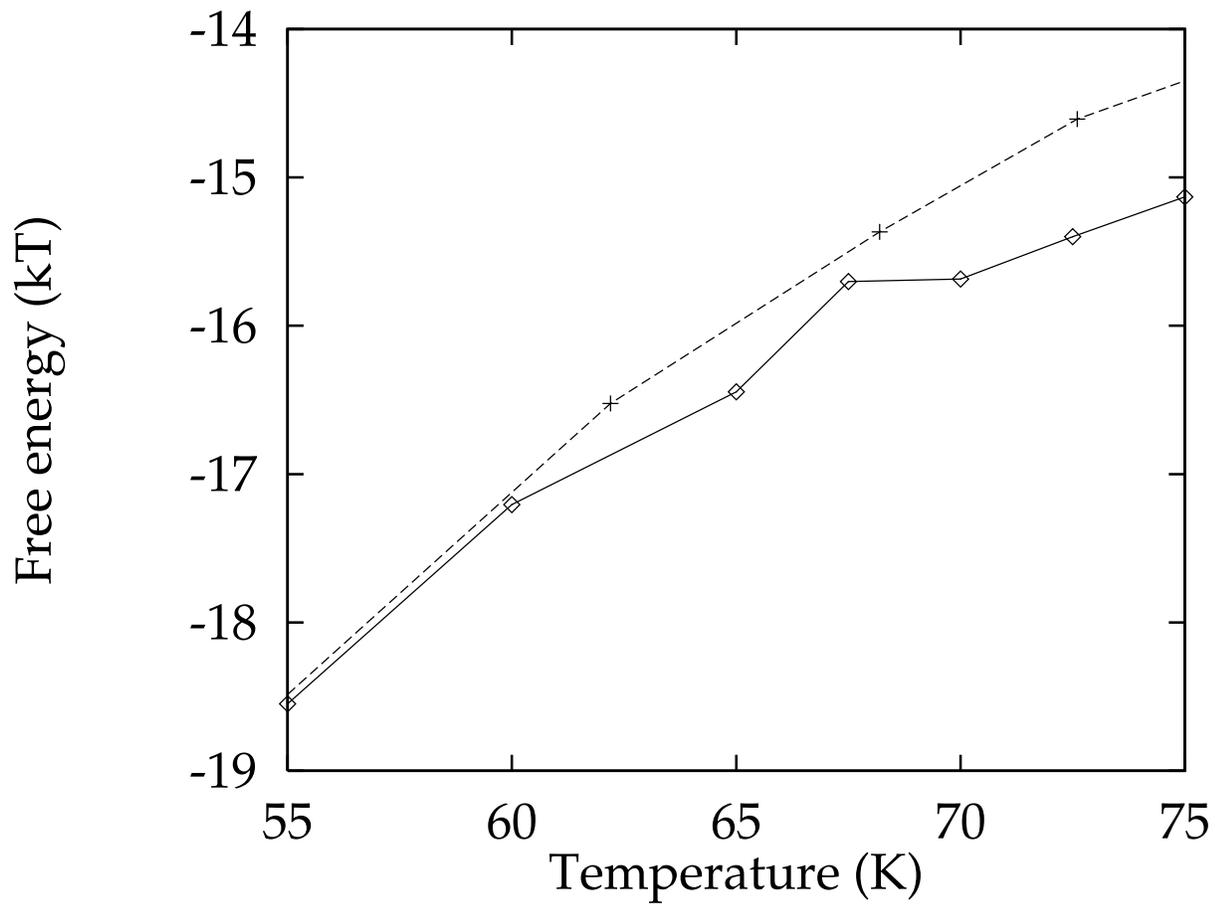
\begin{figure}[p]
   \caption{Gibbs free energy of solid Argon. The dashed line is
   calculated from the experimentally known vapor pressure, the solid
   line is calculated using the transfer matrix method described in the
   text.}
   \label{gibbs2}
\end{figure}

The kinetic part of Eq.~(\ref{partfunc}) is easily integrated.
Substituting the appropriate values (Argon has a mass of 39.948 atomic
units), gives the Helmholtz free energy which equals the
Gibbs free energy $G$ because $P=0$, so
\begin{equation}
   \frac{\beta G}{N} =
      -\frac{3}{2} \ln \left( \frac{2\pi m}{\beta h^2} \right)
      -\frac{1}{3} \ln \lambda_{\text{max}} .
\end{equation}
The Gibbs free energy per particle equals the chemical potential $\mu$. The
vapor, being very sparse, behaves as an ideal gas and has a chemical
potential $\mu$,
\begin{equation}
   \beta\mu = \ln ( \lambda^3 \beta P ) ,
\end{equation}
where $P$ is the vapor pressure and $\lambda$ the thermal wavelength,
\begin{equation}
   \lambda = \sqrt{\frac{h^2\beta}{2\pi m}}.
\end{equation}
We plot the Gibbs free energy together with the experimental
values,\cite{handbook} calculated from the vapor pressure, in
Fig.~\ref{gibbs2}. Regarding the fact that all approximations are present
in the plot, the correspondence is remarkable.

\subsection{Cluster calculations}

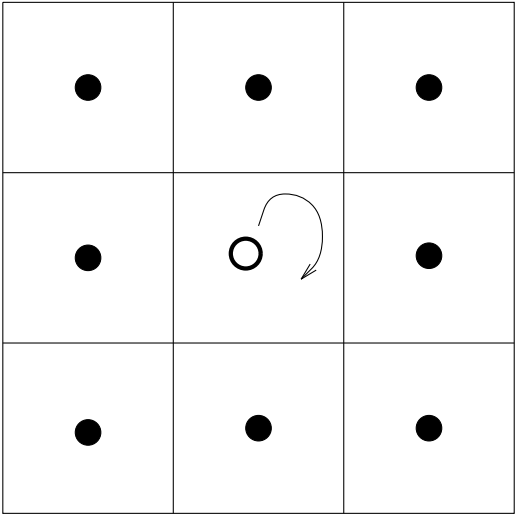
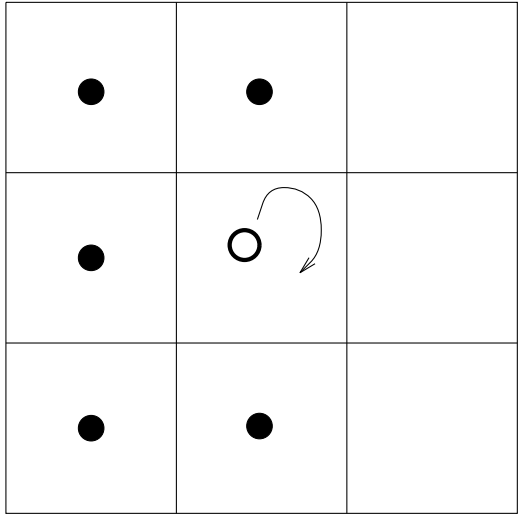
\begin{figure}[p]
   \caption{Two surface configurations used for a cluster calculation.
   Symbols:
   {\protect\large $\bullet$} are atoms fixed in the center of their cells,
   {\protect\large $\circ$} are the atoms that move. (a) is a fully occupied
   surface, (b) is a surface with a step. The rightmost cells of (b)
   are empty.}
   \label{cluster}
\end{figure}

We will now check the error arising from going from a three dimensional,
dense grid of points to a two dimensional grid consisting of 25 points
in the cells. We use a cluster calculation of a small configuration of
particles at the surface. In this configuration, we fix all particles at
their equilibrium positions, except for one. This particle is moving
through its cell in the field of the others. We choose the two
configurations depicted in Fig.~\ref{cluster}; in Fig.~\ref{cluster}(a)
all cells are occupied, in Fig.~\ref{cluster}(b) there is a step present.
The central particle in both of the figures is the particle that moves.

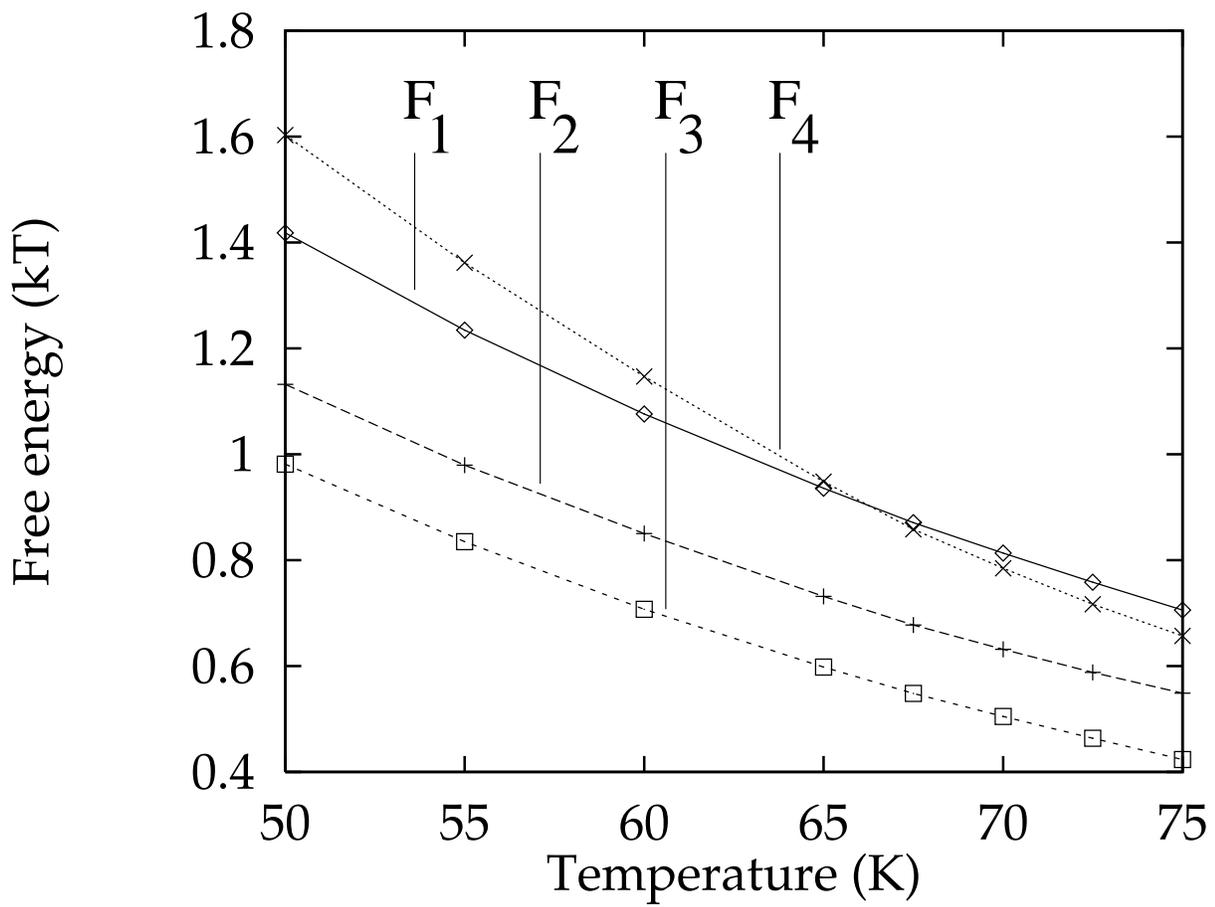
\begin{figure}[p]
   \caption{Free energies $F_1$ to $F_4$ of the four different vertex
   configurations of Fig.~\protect\ref{vertices}. Symbols:
   $\Diamond$~straight step, $+$~inside corner, $\Box$~outside corner,
   $\times$~crossing. The temperature dependence of the crossing of two
   steps is the strongest, which means that there the entropy plays a more
   important role.}
   \label{freeenergies}
\end{figure}

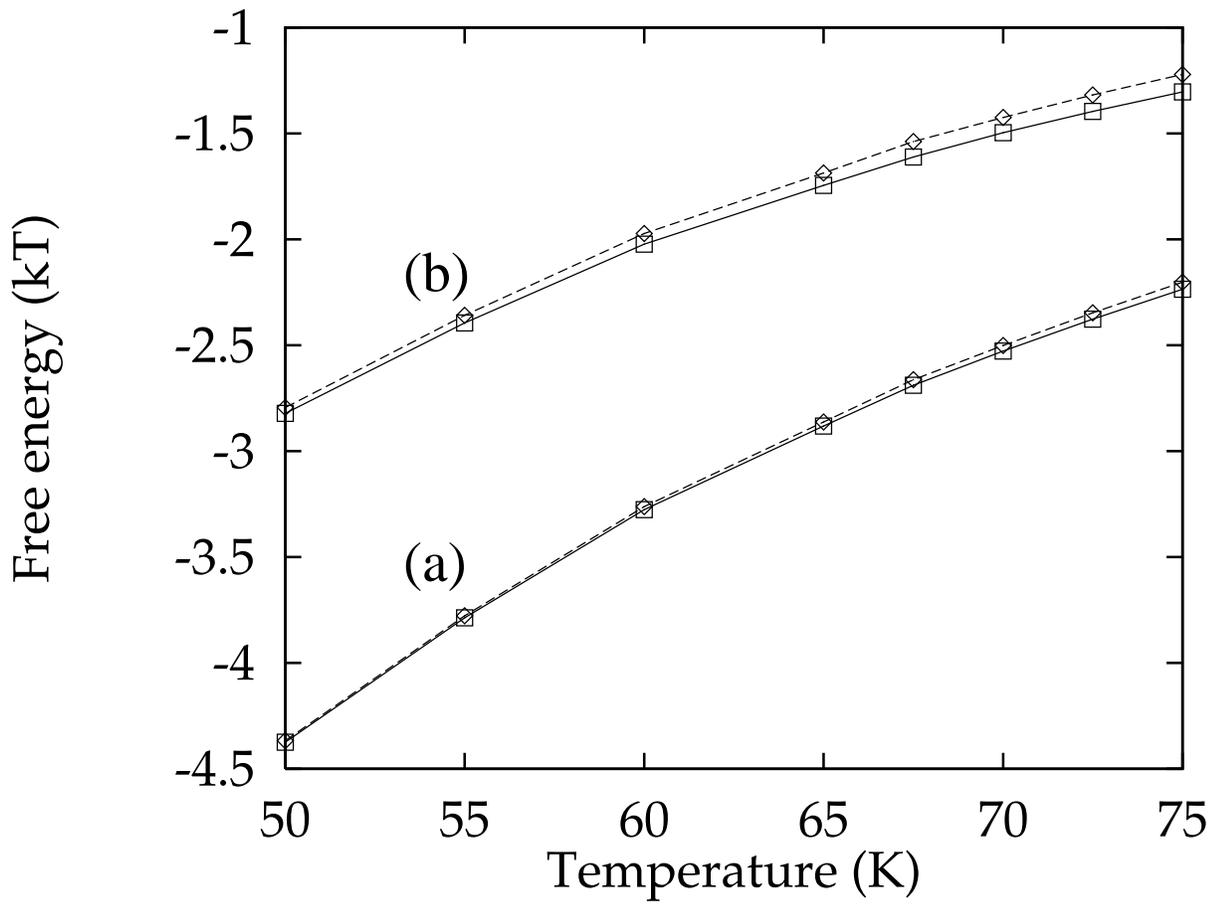
\begin{figure}[p]
   \caption{Free energy of the clusters depicted in
   Fig.~\protect\ref{cluster}, calculated in two different ways. (a)
   free energy of the fully occupied configuration from
   Fig.~\protect\ref{cluster}(a), (b) is from the
   configuration with a step in Fig.~\protect\ref{cluster}(b). Symbols:
   solid lines and $\Box$ are calculated using the full three
   dimensional potential pattern, dashed lines and $\Diamond$ are
   the results after averaging the potential pattern over its vertical
   coordinate.}
   \label{clusterfig}
\end{figure}

We calculate the free energy of both configurations in two ways: the first
using the full three dimensional potential pattern consisting of
$21 \times 21 \times 21$ points. In the second calculation, the pattern is
first averaged over its vertical coordinate, thus consisting of
$21 \times 21$ points. To obtain an indication of the step free energy, we
divide the potential by two, as it should in a cluster calculation. Results
are plotted in Fig.~\ref{clusterfig}, showing that the averaging hardly
has any effect. Note that the difference between the two plots (a) and
(b) indicates the step free energy. Compare this with the step free
energies plotted in Fig.~\ref{freeenergies}, notably the values for
$F_1$. It follows that a cluster calculation still overestimates the
step energy.

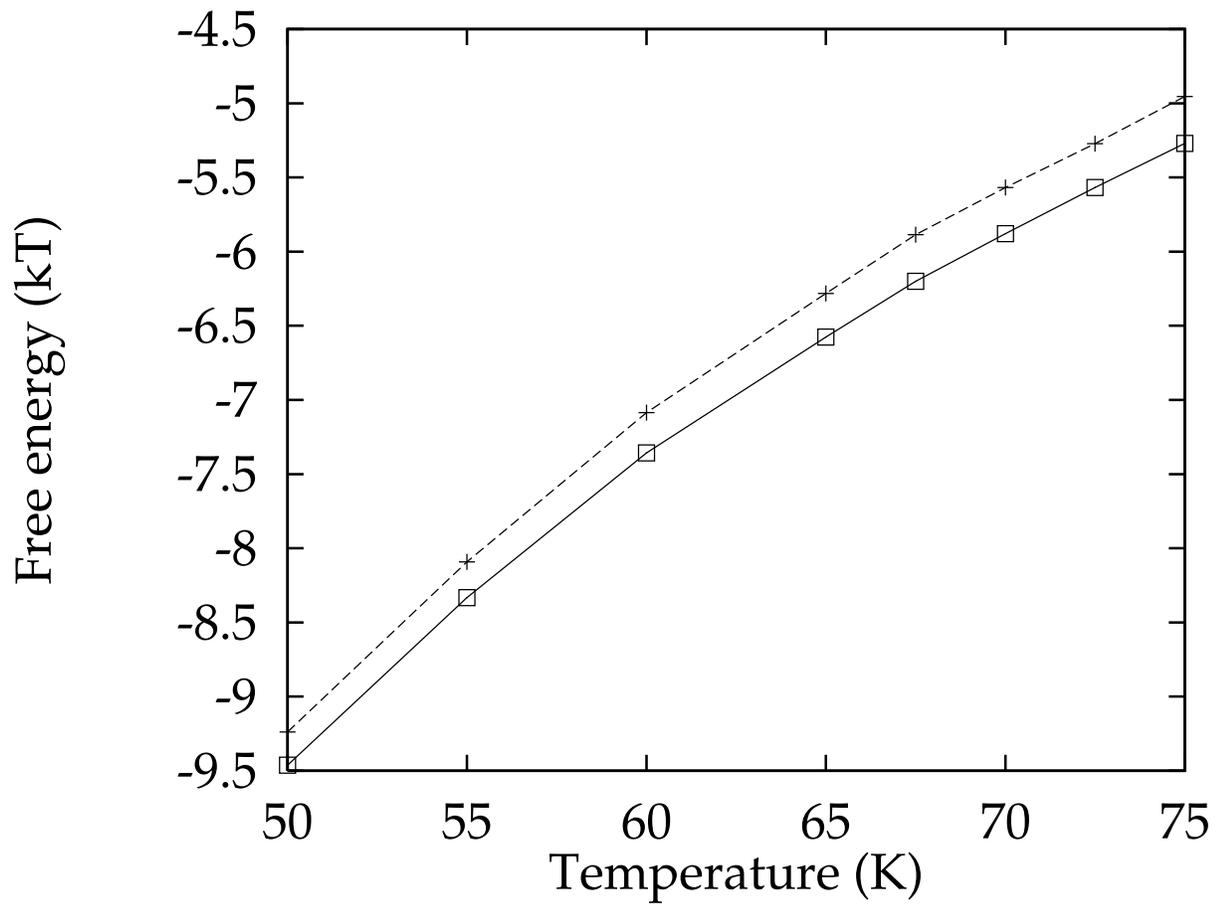
\begin{figure}[p]
   \caption{Free energy of the cluster depicted in
   Fig.~\protect\ref{cluster}(a). Solid lines with $\Box$ are the exact
   results, using the potential pattern defined on a fine grid of $21
   \times 21 \times 21$ points. The dashed line with $+$ is the
   result from a transfer matrix calculation, using a two dimensional
   grid consisting of 25 points.}
   \label{clustrans}
\end{figure}

Next we test the effect of sparsing the grid. We calculate the
free energy of the configuration in Fig.~\ref{cluster}(a), but now using
the transfer matrix with the grid consisting of 25 points. For comparison,
we plot the results together with the exact calculation. The plot is
shown in Fig.~\ref{clustrans}, showing that sparsing the grid is a
reasonable approximation. Note that we used the full potential in
this calculation, whereas the calculation from Fig.~\ref{clusterfig} is
done with the potential divided by two.
\\[2ex]
We conclude, from both checks, that within a reasonable accuracy, the
approximations used are valid.

\begin{table}
   \caption{Comparison of the vertex free energies $F_j$ in units of
   $kT$ at $T=50$~K and
   $T=70$~K. The second and third columns show the actual free energies
   of the vertices 1 to 4. In the fourth column, the same free energies
   are calculated, but now using the potential pattern at $T=50$~K
   scaled to the cell dimensions of $T=70$~K. The last column shows the
   relative contribution of the increased cell dimensions to the total
   effect. It follows that the decrease of free energy with increasing
   temperature is for roughly 98\% due to the increasing cell dimensions,
   and for 2\% to the flattening potential pattern.}
   \label{checktable}
   \begin{tabular}{ccccc}
      Vertex number & 50 K & 70 K actual & 70 K scaled & \% \\
      \tableline
      1 & 1.4180 & 0.8133 & 0.8219 & 98.6 \\
      2 & 1.1322 & 0.6314 & 0.6390 & 98.5 \\
      3 & 0.9814 & 0.5048 & 0.5130 & 98.3 \\
      4 & 1.6028 & 0.7848 & 0.8018 & 97.9 \\
   \end{tabular}
\end{table}

%

\begin{figure}
   \caption{Nine surface configurations. The squares are the cells in
   which the surface atoms move, empty squares denoting empty cells and
   shaded squares denoting occupied cells. Each of the surface
   configurations is understood to extend to infinity on both sides.
   Steps on the surface (that is, borders between empty and occupied
   cells) make up closed oriented loops, and are indicated with thick
   lines.}
   \label{confs}
\end{figure}

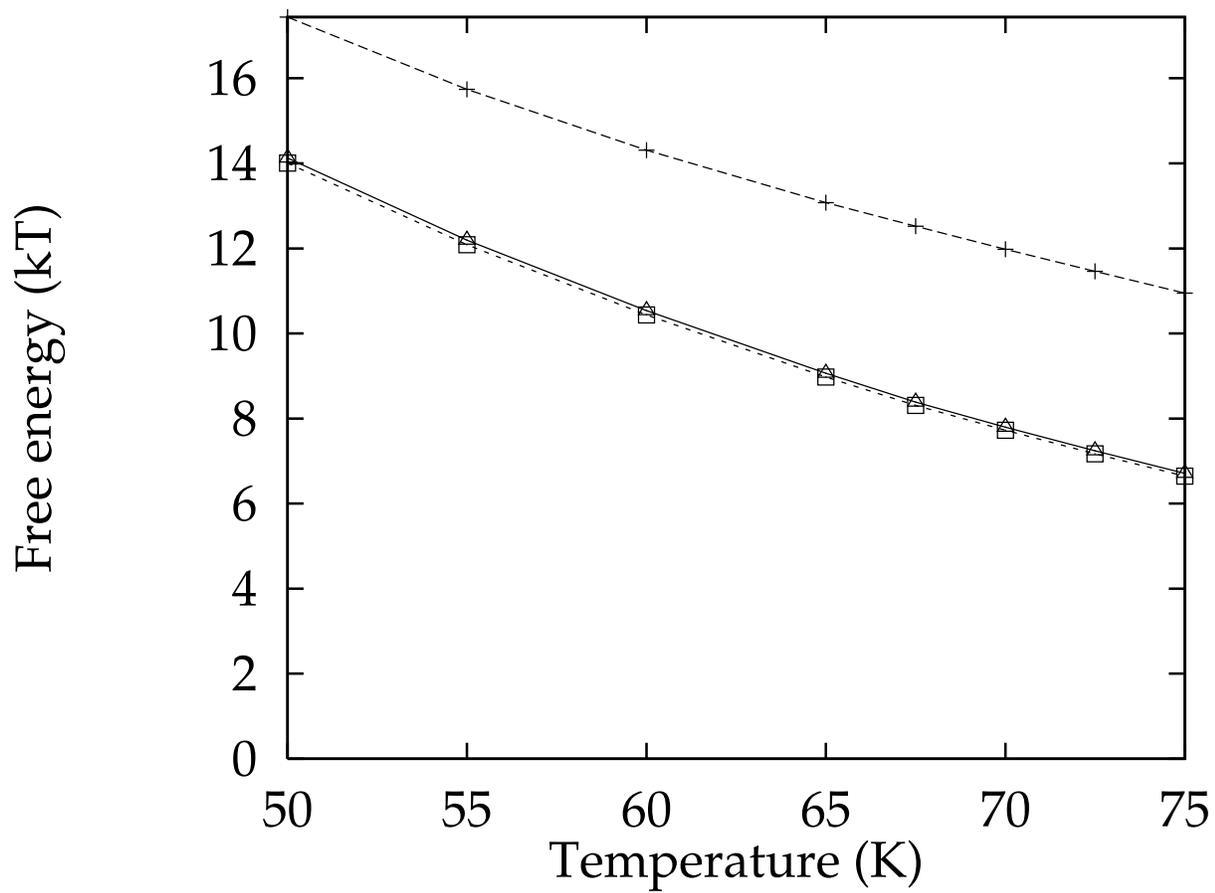
\begin{figure}[p]
   \caption{Free energy of surface configuration number 7 from
   Fig.~\protect\ref{confs}. Symbols: $\Box$~is the calculated free energy,
   $+$~is the energy in terms of broken bonds, and $\triangle$~is the
   free energy calculated from the vertex free energies $F_j$ from
   Fig.~\protect\ref{vertices}. It follows that expressing the free energy
   of the surface in terms of the vertex free energies is accurate; $\Box$
   and $\triangle$ are approximately the same. Also follows that
   calculating the energy of a surface configuration by counting the
   broken bonds considerably overestimates the actual value; the figure
   shows that a substantial entropy effect is present. As expected, this
   effect gets stronger with increasing temperature.}
   \label{conf7}
\end{figure}

\end{document}